\documentclass[twocolumn,prl, superscriptaddress, nopacs]{revtex4}
\newcommand{\pagenumbaa}{1}
\usepackage{graphicx}
\usepackage{amssymb}
\usepackage{amstext}
\usepackage{epsfig}

\usepackage{latexsym}
\usepackage{amsmath}
\usepackage{bm} 
\usepackage{bbm}

\usepackage{color}

\usepackage{amsthm}
\usepackage{amsmath}
\usepackage{amsfonts}
\usepackage{amssymb,amstext}
\usepackage{appendix}
\usepackage{enumitem}

\theoremstyle{plain}

\theoremstyle{definition}

\usepackage{epsfig}
\usepackage{algpseudocode}
\usepackage{amscd}
\usepackage{algorithm}
\usepackage{tikz}

\newcommand{\beq}{\begin{equation}}
\newcommand{\eeq}{\end{equation}}

\newcommand{\beqa}{\begin{eqnarray}}
\newcommand{\eeqa}{\end{eqnarray}}

\newcommand{\bal}{\begin{align}}
\newcommand{\eal}{\end{align}}

\newcommand{\bsp}{\begin{equation}\begin{split}}
\newcommand{\esp}{\end{split}\end{equation}}

\newcommand{\bit}{\begin{itemize}}
\newcommand{\eit}{\end{itemize}}

\newcommand{\ben}{\begin{enumerate}}
\newcommand{\een}{\end{enumerate}}

\newcommand{\nn}{\nonumber}
\newcommand{\SPAN}{\text{span}}

\renewcommand{\sp}[2]{\langle #1,#2 \rangle}
\newcommand{\AR}{\mathbb{R}}

\newcommand*{\ket}[1]{| #1 \rangle}

\newcommand{\HR}{\mathcal{H}}
\newcommand{\ketbra}[1]{| #1 \rangle \langle #1 |}

\newcommand{\id}{\mathbb{I} }

\newcommand{\tr}{\mathrm{tr}}
\newcommand{\rank}{\mathrm{rank}}

\newcommand{\Herm}{\mathrm{Herm}}
\newcommand{\data}{\mathcal{D}}

\newcommand{\knowledge}{\vec{\mathcal{K}}}

\begin{document}


\title{Self-consistent tomography of the state-measurement Gram matrix}


\author{Cyril Stark}

\affiliation
{Theoretische Physik, ETH Zurich, CH-8093 Zurich, Switzerland}


\begin{abstract}

State and measurement tomography make assumptions about the experimental states or measurements. These assumptions are often not justified because state preparation and measurement errors are unavoidable in practice. Here we describe how the Gram matrix associated with the states and measurement operators can be estimated via semidefinite programming if the states and the measurements are so called globally completable. This is for instance the case if the unknown measurements are known to be projective and non-degenerate. The computed Gram matrix determines the states, and the measurement operators uniquely up to simultaneous rotations in the space of Hermitian matrices. We prove the reliability of the proposed method in the limit of a large number of independent measurement repetitions.

\end{abstract}


\maketitle

\setcounter{page}{\pagenumbaa}
\thispagestyle{plain}


We consider an experiment which allows the preparation of $W$ different states $(\rho_{w})_{w=1}^W$ and the performance of $V$ different measurements, each described in terms of measurement operators $(E_{vk})_{k=1}^K$. Repeating the measurement of $\rho_{w}$ with $(E_{vk})_{k=1}^K$ $N$ times, we can count how many times we have measured the outcomes ``$1$'', ..., ``$K$''. Dividing these numbers by $N$ we obtain frequencies $f_{k|wv}$ for measuring ``$k$'' given that we have prepared the state ``$w$'' and performed the measurement ``$v$''. The data table
\beq\label{def.data.D}
	\mathcal{D} =   \left( \begin{array}{ccccccc}  f_{1|11} & \cdots & f_{K|11} & \cdots &   f_{1|1V} & \cdots & f_{K|1V}  \\  f_{1|21} & \cdots & f_{K|21} & \cdots &   f_{1|2V} & \cdots & f_{K|2V} \\  \vdots &   & \vdots &   &   \vdots &   & \vdots  \\   f_{1|W1} & \cdots & f_{K|W1} & \cdots &   f_{1|WV} & \cdots & f_{K|WV}  \end{array} \right) \nn
\eeq
describes the phenomenological content of the experiment. Changing the row index in $\data$ amounts to changing the state while changing the column index in $\data$ amounts to changing the measurement outcome. In the limit $N \rightarrow \infty$, by Born's rule,
\beq\label{Born.rule}
	f_{k|wv} = \tr( \rho_{w} E_{vk} ).
\eeq
The matrices $\rho_{w}$ and $E_{vk}$ are contained in the space of Hermitian matrices $\Herm(\mathbb{C}^d)$ if the underlying quantum system is $d$-dimensional. With respect to an othonormal basis in $\Herm(\mathbb{C}^d)$ we can express them in terms of vectors $\vec{\rho}_{w}, \vec{E}_{vk} \in \AR^{d^2}$ because $\Herm(\mathbb{C}^d)$ is a real $d^2$-dimensional vector space. It follows that
\beq\nn
	\tr( \rho_{w} E_{vk} ) = (\vec{\rho}_{w})^T \vec{E}_{vk}.
\eeq
Define the matrices
\beq\label{def.matrix.carrying.st.and.m}
\begin{aligned}
	P_{\mathrm{st}} &:= (\vec{\rho}_{1} \, | \cdots | \, \vec{\rho}_{W}),\\
	P_{\mathrm{m}} &:= ( \vec{E}_{11}  \, | \cdots | \, \vec{E}_{1K} \, | \cdots  | \, \vec{E}_{V1} \, | \cdots | \, \vec{E}_{VK} ),\\
	P &:= ( P_{\mathrm{st}} \; | \; P_{\mathrm{m}}  ),
\end{aligned}	
\eeq
and $G = P^{T}P$. The matrix $G$ is the Gram matrix associated with $\rho_{1}$, ..., $E_{VK}$. The data table $\mathcal{D}$ appears as off-diagonal block in the states-measurement Gram matrix: 
\beq\label{data.as.block.of.Gram}
	G = \left( \begin{array}{c|c}  G_{\mathrm{st}} & \mathcal{D} \\ \hline  \mathcal{D}^T & G_{\mathrm{m}}   \end{array} \right)
\eeq
Our goal is to determine $G$. Closely related is the work~\cite{merkel2012self} by Merkel and coworkers. They introduce a semidefinite program to perform process tomography. Their method is not prone to state preparations and measurement errors if the experimental gates are well approximated by the ideal gates (i.e., the gates we would like to be realized experimentally). On the other hand, Kimmel~\emph{et al.} were able to show in their seminal work~\cite{kimmel2013robust} that randomized benchmarking can be used to estimate the unital part of trace-preserving quantum operations. The branch of quantum tomography aiming at the simultaneous estimation of the underlying states and measurements has recently been termed self-consistent state and measurement tomography~\cite{medford2013self,takahashi2013tomography}. Hence, self-callibrating quantum tomography~\cite{MogilevtsevHradil2012,MogilevtsevHradil2009,Mogilevtsev2010PRA,James2012NJP} and self-consistent state and measurement tomography share the same goals. We would like to point out the relation between this paper and dimension estimation~\cite{Christandl2008PRA,Brunner2008PRL,Vertesi2009PRA,Wolf2009PRL,Brunner2010PRL,GallergoAcin2012}, the detection of temporal drifts~\cite{FlammiaGross2012}, the falsification of ansatz models~\cite{MoroderBlattSystematicErrors2012} and the analysis of the implications of defective assumptions (i.e., systematic errors) about the considered experiment~\cite{AcinGisinMasanes2006,Scarani2011FalseBellViolation,RossetGisin2012}.

In the analysis of our method we neglect statistical fluctuations but we take into account systematic errors. Hence, the results of this paper are orthogonal to the recent constructions of confidence regions~\cite{Robin2011,ChristandlRenner2012} for state estimation which take into account statistical fluctuations but neglect systematic errors. 

\emph{Gram estimation.} Let $\Omega$ be the set of indices marking the known entries of $G$, and let $\vec{\mathcal{K}} \in \AR^{| \Omega |}$ be the explicit numerical values of the known entries. Thus, $G_{\Omega} = \vec{\mathcal{K}}$ captures our a priori knowledge about the entries of $G$. In the remainder we assume that at least the phenomenological data~$\data$ is known and thus part of $\vec{\mathcal{K}}$.

Note that $\rank(G) = \rank(P)$. We can only hope to successfully reconstruct general states and measurements if 
\beq\label{Gram.est:Eq.the.spanning.assumption}
			\Herm(\mathbb{C}^{d}) = \left\{ \begin{array}{l}  \SPAN_{\AR}\{ \rho_{w} \}_{w},   \\    \SPAN_{\AR}\{ E_{vk} \}_{vk}.   \end{array} \right.
\eeq
Otherwise, we could not even determine the states if all measurements were known and vice versa. Hence, both matrices $P_{\mathrm{st}}$ and $P_{\mathrm{m}}$ are assumed to be full-rank in the remainder. Recalling $\mathcal{D} = P_{\mathrm{st}}^T P_{\mathrm{m}}$, it follows that
\beq\label{eq.forming.the.basis.for.the.consistency.test}
	\rank(\mathcal{D}) = \rank(P_{\mathrm{m}}) = \rank(G) = d^2.
\eeq
Let $\mathcal{G}_{\mathrm{QM}}$ denote the set of Gram matrices that can be generated via quantum density matrices and measurement operators. The set $\mathcal{G}_{\mathrm{QM}}$ is contained in the set $S^+(\AR^{N})$ of symmetric and positive semidefinite matrices on $\AR^{N}$ ($N := W+VK$). By Eq.~\eqref{eq.forming.the.basis.for.the.consistency.test}, determining the state-measurement Gram matrix is equivalent to solving the problem
\beq
\begin{aligned}\label{Gram.estimation:original.feasiblity.problem}
	 &\text{find} & & G \in \mathcal{G}_{\mathrm{QM}} \\
	& \text{subject to}
	& & G_{\Omega} = \vec{\mathcal{K}}, \\
	&&& \rank(G) = \rank(\data).
\end{aligned}
\eeq
Since $\data$ is a partial matrix of $G$, $\rank(\data) \leq \rank(G)$. Therefore, instead of solving~\eqref{Gram.estimation:original.feasiblity.problem} we can equally well solve the optimization problem~\footnote{Hence, our method is reminiscent of compressed sensing~\cite{Gross2010,FlammiaGross2012}.}
\beq
\begin{aligned}\label{Gram.estimation:original.feasiblity.problem.as.rank.min}
	 &\text{minimize} & & \rank(G) \\
	& \text{subject to}
	& & G_{\Omega} = \vec{\mathcal{K}}, \\
	&&& G \in \mathcal{G}_{\mathrm{QM}}.
\end{aligned}
\eeq
When trying to compute~\eqref{Gram.estimation:original.feasiblity.problem.as.rank.min} we face two major difficulties: 

\begin{description}[style=multiline,leftmargin=0.8cm,font=\normalfont]
\item[(D1)]	The optimization problem~\eqref{Gram.estimation:original.feasiblity.problem.as.rank.min} is not convex (thus leading to local minima~\footnote{Rank minimization is even NP hard.~\cite{Recht2010}}) because the rank of a matrix is not a convex function. For example,
\beqa\nn
	2 
	&=&	\rank \bigl( p \; \ketbra{0} + (1-p) \; \ketbra{1} \bigr) \nn \\
	&>&	p \; \rank\bigl(\ketbra{0}\bigr) + (1-p) \; \rank\bigl(\ketbra{1}\bigr) = 1.\nn
\eeqa
\item[(D2)]	We do not know how to efficiently characterize the set of quantum Gram matrices $\mathcal{G}_{\mathrm{QM}}$.
\end{description}

\begin{figure}[tbp]
\centering
\includegraphics[width=0.8\columnwidth]{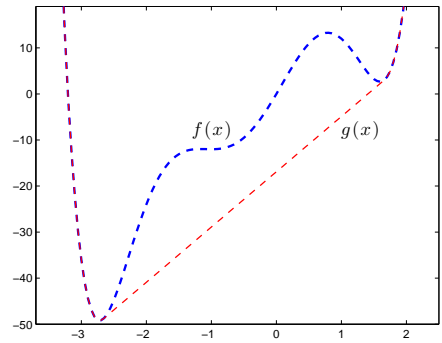}
\caption{The function $g(x)$ is the convex envelope of $f(x)$, i.e., it is the largest convex function that pointwise lower bounds $f(x)$.}
\label{fig:figuresconv_env_for_illustration}
\end{figure}

\emph{Difficulty (D1).} One approach to solve non-convex optimization problems is to relax them to a closely related convex optimization problem. This is illustrated in Fig.~\ref{fig:figuresconv_env_for_illustration}: instead of trying to find the global minimum of the non-convex function $f(x)$, we are computing the minimum of the function $g(x)$. Then, the convexity of $g(x)$ guarantees that the found minimum is a global minimum of $g(x)$. A function $g: \; \mathcal{C} \rightarrow \AR$ from a convex set $\mathcal{C}$ to $\AR$ is the \emph{convex envelope} of a function  $f: \; \mathcal{C} \rightarrow \AR$ if it is the pointwise largest convex function satisfying $g(x) \leq f(x)$ for all $x \in \mathcal{C}$. Note that this property depends on the convex set $\mathcal{C}$. This becomes evident when we replace the choice $\mathcal{C}=\AR$ in Fig.~\ref{fig:figuresconv_env_for_illustration} with an interval, e.g., to $\mathcal{C} = [-1,1]$. Recall that $\mathcal{G}_{\mathrm{QM}} \subset S^+(\AR^N)$. Our goal is to solve the optimization problem~\eqref{Gram.estimation:original.feasiblity.problem.as.rank.min} by replacing the rank function by its convex envelope with respect to the convex set
\beq\nn
	\mathcal{C} := \{ X \in \AR^{n\times n} \; |Ê\; X \geq 0, \; \| X \| \leq R_{\mathrm{QM}}(d) \},
\eeq	
where $\| . \|$ denotes the operator norm and
\beq\label{Gram.est:Def.M.QM}
	R_{\mathrm{QM}}(d) := \sup\{ \| X \| \; | \; X \in \mathcal{G}_{\mathrm{QM}}, \; \dim(\HR) \leq d \},
\eeq
i.e., $R_{\mathrm{QM}}(d)$ is the radius of the smallest operator norm ball containing the Gram matrices corresponding to $d$-dimensional quantum systems. In the supplementary material we show that  $R_{\mathrm{QM}}(d) = W + Vd$. Fazel, Hindi and Boyd proved~\cite{FazelHindiBoyd2001} that $\| X \|_{1} / R_{\mathrm{QM}}$ is the convex envelope of the rank function on the larger set 
\beq\nn
	\mathcal{C}' := \{ X \in \AR^{n\times n} \; |Ê\; \| X \| \leq R_{\mathrm{QM}}(d) \} \supset \mathcal{C}.
\eeq 
Here, $\| . \|_{1}$ denotes the trace norm. Since $\mathcal{C}$ is a proper subset of $\mathcal{C}'$ it is unclear whether or not $\| X \|_{1} / R_{\mathrm{QM}}$ is also the convex envelope of the rank function with respect to $\mathcal{C}$ (see Fig.~\ref{fig:figuresconv_env_for_illustration}). That this is indeed the case is proven in the supplementary material by adapting the derivation~\cite{FazelHindiBoyd2001} given by Fazel, Hindi and Boyd. Consequently, we arrive at the following substitution of the non-convex optimization problem~\eqref{Gram.estimation:original.feasiblity.problem.as.rank.min}:
\beq\label{trace.minimization.draft}
\begin{aligned}
        & \text{minimize} & & \tr \, G \\
	& \text{subject to}
	& & G_{\Omega} = \knowledge, \\
	&&&G \in S^+(\AR^N), \; \|G\| \leq R_{\mathrm{QM}}(d).
\end{aligned}
\eeq
The optimization problem~\eqref{trace.minimization.draft} can be cast into a semidefinite program (SDP): the constraint $\|G\| \leq R_{\mathrm{QM}}$ in Eq.~\eqref{trace.minimization.draft} is equivalent to $R_{\mathrm{QM}}\id - G\geq 0$ because $G \geq 0$. Hence, the optimization problem~\eqref{trace.minimization.draft} is equivalent to the SDP
\beq\label{trace.minimization}
\begin{aligned}
        & \text{minimize} & & \tr \, G \\
	& \text{subject to}
	& & G_{\Omega} = \knowledge, \\
	 & & & Z =  ( W + Vd ) \id -G \\ 
	 & & & G, Z \geq 0,
\end{aligned}
\eeq
because
\beq
\begin{split}\nn
	Z \geq 0, \; G + Z = M_{QM}(d)\id, \; \Rightarrow M_{QM}(d)\id - G \geq 0.
\end{split}	
\eeq
Consequently, the optimization problem~\eqref{trace.minimization.draft} can be solved reliably and efficiently by standard methods~\cite{Sturm1999, BeckerCandesGrant2010}.

\emph{Difficulty (D2).} Even though closely related, the problems~\eqref{Gram.estimation:original.feasiblity.problem.as.rank.min} and~\eqref{trace.minimization.draft} are not identical. There exists no guarantee that the global optimum of the semidefinite program~\eqref{trace.minimization.draft} and the global optimum of the rank minimization~\eqref{Gram.estimation:original.feasiblity.problem.as.rank.min} agree. In particular, when going from the original rank minimization to its convex relaxation, we extended the feasible set by replacing $\mathcal{G}_{\mathrm{QM}}$ with $\{ X \geq 0, \; \| X \| \leq R_{\mathrm{QM}}(d) \}$. Hence, there is no guarantee that the solution of the relaxed problem~\eqref{trace.minimization.draft} lies in~$\mathcal{G}_{\mathrm{QM}}$. 

However, if there existed only one Gram matrix $G$ satisfying $G_{\Omega} = \vec{\mathcal{K}}$ and $\rank(G) = \rank(\data)$ and if the solution $\hat{G}$ of~\eqref{trace.minimization} satisfies $G_{\Omega} = \vec{\mathcal{K}}$ and $\rank(G) = \rank(\data)$, then we know that $\hat{G}$ is equal to the correct state-measurement Gram matrix. 

Is the state-measurement Gram matrix $G$ ever uniquely determined in this way? Generically~\cite{Stark2013UniquenessInSelfConsistentTomography} $G$ is never uniquely fixed if $\data$ is all we know about $G$. Hence, we need to make sure that our knowledge $G_{\Omega} = \vec{\mathcal{K}}$ and $\rank(G) = \rank(\data)$ is sufficient to uniquely determine $G$. In~\cite{Stark2013RigidityOfGramMatrices}, a necessary and sufficient criterium for the uniqueness of $G$---given $G_{\Omega} = \vec{\mathcal{K}}$ and $\rank(G) = \rank(\data)$---is derived for situations in which either something is known about $G_{\mathrm{st}}$ or $G_{\mathrm{m}}$. Then, $G$ is uniquely determined if and only if
\[
	\rank\left(\mathcal{M}(\vec{\mathcal{K}}, \Omega)\right) = d^2 (d^2 + 1)/2
\]
where $\mathcal{M}(\cdot)$ is a matrix-valued function in $\vec{\mathcal{K}}$ and $\Omega$ (see~\cite{Stark2013RigidityOfGramMatrices}). Applying this criterion for specific circumstances we observe (for instance) that knowing that the experimental measurements are projective and non-degenerate is sufficient to ensure uniqueness of $G$ if the number of measurements exceeds a critical value.

\emph{Algorithm.} The previous discussions lead to Algorithm~\ref{procedure.including.consistency.test} for estimating the state-measurement Gram matrix. 

\begin{algorithm}
\caption{Gram estimation}
\label{procedure.including.consistency.test}
\begin{algorithmic}[1]
\Require $G \geq 0$ is determined uniquely by $G_{\Omega} = \vec{\mathcal{K}}$ and $\rank(G) = \rank(\data)$~\cite{Stark2013RigidityOfGramMatrices}.
	
	\State Run the optimization~\eqref{trace.minimization.draft} to compute a completion $\hat{G}$ of $G_{\Omega}$.
	
	\While{$\rank(\hat{G}) > \rank(\mathcal{D})$}
		\State Prepare additional states, or perform additional 
		\State measurements.
		\State Run the optimization~\eqref{trace.minimization.draft}.
	\EndWhile		
\State Return $\hat{G}$.	
\end{algorithmic}
\end{algorithm}

\emph{Numerical examples.} We are going to run Algorithm~\ref{procedure.including.consistency.test} for different Hilbert space dimensions $d$ assuming that we know that the performed measurements are projective and non-degenarate. We will observe that the estimations $\hat{G}$ frequently fail to satisfy $\rank(\hat{G}) = \rank(\mathcal{D})$ whenever $\Omega$, $d$, $W$, $V$ and $K$ are such that $G$ is not overdetermined by $G_{\Omega}$. 

The results are summarized in Table~\ref{numerical.experiments}. The second and the third columns list the number of successful respectively failed reconstructions of the full Gram matrix $G$. Here, 
\beq
	\text{`failure'} \; :\Leftrightarrow \; \max_{ij} | \hat{G}_{ij} - G_{ij} | \geq 10^{-3},
\eeq
with $\hat{G}$ and $G$ denoting the estimated and the correct Gram matrix respectively. 'Start point' refers to the number of states and measurements we use to start Algorithm~\ref{procedure.including.consistency.test}. These start points are chosen such that $P$ is globally completable. If the trace minimization~\eqref{trace.minimization.draft} fails to satisfy $\rank(\hat{G}) = \rank(\mathcal{D})$, we alternatingly add a new state or a new measurement and re-run the trace minimization~\eqref{trace.minimization.draft}; cf. Algorithm~\ref{procedure.including.consistency.test}. More detailed information about these numerical experiments can be found in the supplementary information.

\begin{table}[htdp]
\caption{Numerical experiments}
\begin{center}
\begin{tabular}{|c||c|c|c|c|}
	\hline   $\dim(\HR)$ & successes & failures & start point & solver  \\ 
	\hline 
	\hline 2  & 2383 & 0 & (5,5) & SeDuMi~\cite{Sturm1999}\\
	 \hline 3  & 1000 & 0 & (60,100) & TFOCS~\cite{BeckerCandesGrant2010}\\
	  \hline 4  & 1000 & 0 & (65,130) & TFOCS~\cite{BeckerCandesGrant2010}\\ 
	\hline
\end{tabular}
\end{center}
\label{numerical.experiments}
\end{table}%

\emph{Discussion.} For Algorithm~\ref{procedure.including.consistency.test} to work provably, we need to assume the following:
\begin{itemize}
\item		\emph{Asymptotic limit.} The number of measurement repetitions of each paring $(\rho_{w},E_{vk})$ is unbounded, so that $f_{k|wv} = \tr( \rho_{w} E_{vk} )$.
\item		\emph{Informationally complete states and measurements.} The unknown matrices $(\rho_{w})_{w}$ and $(E_{vk})_{vk}$ separately span the space of Hermitian matrices associated to   the underlying Hilbert space $\HR$, i.e.,
		\beq
			\Herm(\HR) = \left\{ \begin{array}{l}  \SPAN_{\AR}\{ \rho_{w} \}_{w},   \\    \SPAN_{\AR}\{ E_{vk} \}_{vk}.   \end{array} \right.
		\eeq 
\item		\emph{Uniqueness.} Our a priori knowledge $G_{\Omega} = \vec{\mathcal{K}}$ implies uniqueness of the Gram matrix $G$ satisfying $\rank(G) = \rank(\data)$~\cite{Stark2013RigidityOfGramMatrices} (see also~\cite{SingerCucuringu,kiraly2012algebraic}).
\end{itemize}
Note that these assumptions are non-exotic and strictly weaker than the assumptions made in asymptotic state tomography and measurement tomography. For instance, in case of state tomography all measurement operators are assumed to be known and consequently, all of the measurement Gram matrix $G_{\mathrm{m}}$ is known a priori. For global completability of the states and the measurements on the other hand it suffices to know (for example) that the measurements we perform are projective with known degeneracy. Such properties are easily enforced by only fixing a block-diagonal of $G_{\mathrm{m}}$; see~\cite{Stark2013RigidityOfGramMatrices}.

The proposed method for computing $G$ suffices to determine the states and the measurements up to simultaneous rotations in $\Herm(\HR)$. Explicit density matrices and measurement operators can be found using a heuristic algorithm~\cite{Stark2012FindingExplicitRealizations} that takes $G$ as input. 

To deal with finite measurement repetitions, we would need to weaken the sharp constraints $f_{k|wv} = \tr( \rho_{w} E_{vk} )$ (see assumption ``Asymptotic limit''), e.g., by enforcing $\tr( \rho_{w} E_{vk} ) \in [ f_{k|wv} - \varepsilon, f_{k|wv} + \varepsilon ]$ for some $\varepsilon > 0$. Assume that under these constraints Algorithm~\ref{procedure.including.consistency.test} returns $\hat{G}$ satisfying $\rank(\hat{G}) \leq \Herm(\HR) = d^2$ where $d$ denotes the dimension of the underlying Hilbert space (assumed to be known). Then, as long as $\varepsilon \ll 1$, we suspect that $\hat{G}$ approximates the correct Gram matrix $G$ well if the states and the measurements are somewhat spread out in the cone of positive semidefinite matrices. It is an important open question to investigate the validity of this intuition and to find convenient upper bounds for the distance between $\hat{G}$ and $G$.

\emph{Acknowledgment.} I would like to thank Johan {\AA}berg for his continuous support. I wish to express my gratitude to Matthias Baur, Matthias Christandl, Dejan Dukaric, Philippe Faist, Hamza and Omar Fawzi, David Gross, Patrick Pletscher, Renato Renner, Lars Steffen, David Sutter, L\'idia del Rio, Michael Walter, Sharon Wulff, and M\'ario Ziman for fruitful discussions. I acknowledge support from the Swiss National Science Foundation through the National Centre of Competence in Research ``Quantum Science and Technology''.

\section{Supplementary information}

\subsection{Computing the bound $R_{\mathrm{QM}}$}\label{Gram.est:Section.Computing.the.bound.M.qm}

The purpose of this section is the computation of the upper bound $R_{\mathrm{QM}}$ from Eq.~(8) in the main text. We start with
\beq\label{bound.op.norm.by.HS.norm}
	\| G \| \leq \| G \|_{2},
\eeq 
where $\| G \|^2_{2} = \tr(G^*G)$, which holds true for general matrices. The bound \eqref{bound.op.norm.by.HS.norm} is tight for $G \in \mathcal{G}_{\mathrm{QM}}$ because for every Hilbert space dimension $d$ and for every rank of $G$, we can choose the vectors $\vec{\rho}_{1}$, ..., $\vec{E}_{VK}$ (the columns of $P$; cf. Eq.~(2) in the main text) such that they are almost parallel to $\AR \id$. Thus, for any choice of $\rank(G) = \rank(P)$, $G$ can become arbitrarily close to a positive semidefinite rank-1 matrix (corresponds to all columns of $P$ being parallel). Thus, the vector of eigenvalues of $G$ becomes arbitrarily close to the vector $(\| G \|, 0, ..., 0)^T$, i.e., $\| G \|$ becomes arbitrarily close to $\| G \|_{2}$. Consequently, the upper bound \eqref{bound.op.norm.by.HS.norm} is tight. We continue by observing that
\beq\label{auxilliary.computation.to.bound.M_qm}
\begin{split}
	\| G \|^2	
	&\leq	\| G \|_{2}^2	=	\| P^TP \|_{2}^2 \leq	\| P \|_{2}^4 \\
	&= 	\left( \sum_{j=1}^{W+VK} \| \vec{P}_{j,\cdot} \|^2_{2} \right)^2\\
	&=	\left(  \sum_{w=1}^W \| \rho_{w} \|_{2}^2    +    \sum_{v=1}^V \sum_{k=1}^K \| E_{vk} \|_{2}^2 \right)^2
\end{split}
\eeq
In the second inequality, we have used the sub-multiplicativity of the Hilbert-Schmidt norm. The Hilbert-Schmidt norm of quantum states is lower bounded by the norm of the maximally mixed state and upper bounded by the norm of pure states. Consequently,
\beq\label{norm.bound.for.states}
	\| \rho_{j} \|_{2} \in \left[ 1/\sqrt{d}, 1 \right] \Rightarrow \| \rho_{j} \|_{2}^2 \leq 1.
\eeq
The condition $\sum_{k} E_{vk} = \id$ implies
\beq
\begin{split}
	d
	&=	\| \id \|_{2}^2	= \Bigl\| \sum_{k} E_{vk} \Bigr\|_{2}^2\\
	&=	\sum_{k} \| E_{vk} \|_{2}^2 + \sum_{k \neq q } \tr( E_{vk} E_{vq} )
\end{split}
\eeq
and therefore,
\beq\label{max.sum.norms.povm.elements}
	\sum_{k} \| E_{vk} \|_{2}^2  \leq d
\eeq
because $\tr( M N) \geq 0$ whenever $M, N \geq 0$. This upper bound is tight because it is achieved by projective, non-degenerate measurements. Using Eq.~\eqref{max.sum.norms.povm.elements} and Eq.~\eqref{norm.bound.for.states} in Eq.~\eqref{auxilliary.computation.to.bound.M_qm}, we arrive at
\beq
	R_{\mathrm{QM}} = W + Vd .
\eeq

\subsection{Finding the convex relaxation with respect to $S^+ \cap B_{\|.\| \leq 1}$}

Fazel, Hindi and Boyd proved~\cite{FazelHindiBoyd2001} that $\| \cdot \|_{1}$ is the convex envelope of the rank function on the set of matrices $X$ with $\| X \| \leq 1$. In the following we present a modification of their argument to show that the trace (and hence still the trace norm $\| \cdot \|_{1}$) is the convex envelope of the rank function when restricting the above ball of matrices $\| X \| \leq 1$ to its intersection with the cone of positive semidefinite matrices, i.e., $X \in S^+ \cap B_{\|.\| \leq 1}$.

Recall that for an arbitrary function $f: \; \mathcal{C} \rightarrow \AR$, $\mathcal{C}$ convex,
\[
	f^*(y) = \sup\{ \sp{y}{x} - f(x) \;Ê| \; x \in \mathcal{C} \}
\] 
is its conjugate. The convex envelope of the rank function with respect to the convex set
\[
	\mathcal{C} := \{ X \in \mathbb{R}^{n \times n} \; | \; X \geq 0, \; \| X \| \leq 1 \}
\]	
is $\rank^{**}$, i.e., the double-conjugate~\cite{HiriartUrruty2} with respect to $\mathcal{C}$. Observe that 
\begin{multline}\label{sup.as.disjoint.rank.sups}
	\rank^*(Y) = \sup_{X \in \mathcal{C}} \{   \tr(YX) - \rank(X)  \} \\
	= \max\Bigl\{    \sup_{\substack{X \in \mathcal{C}, \\ \rank(X)=1}} \{   \tr(YX) - 1  \} , ...,  \sup_{\substack{X \in \mathcal{C}, \\ \rank(X)=n}} \{   \tr(YX) - n  \}     \Bigr\}.
\end{multline}
Here, $Y$ is an arbitrary Hermitian $(n \times n)$ matrix (recall that the Hermitian matrices form the vector space carrying $S^+$). Due to their Hermiticity, both $X$ and $Y$ can be diagonalized orthogonally, 
\beq
\begin{split}
	X &= \sum_{j=1}^{n} \varepsilon(X)_{j} \ketbra{\varepsilon(X)_{j}},\\
	Y &= \sum_{j=1}^{n} \varepsilon(Y)_{j} \ketbra{\varepsilon(Y)_{j}}.
\end{split}
\eeq
In the remainder we are assuming that all the eigenvalues are sorted descendingly. We observe that
\beq
\begin{split}
	\tr(YX) &= \sum_{i=1}^n \varepsilon(Y)_{i} \left( \sum_{j=1}^n \varepsilon(X)_{j} |\langle \varepsilon(X)_{i} | \varepsilon(Y)_{j} \rangle |^2  \right)\\
		   &= \vec{\varepsilon}(Y)^T \, Q \, \vec{\varepsilon}(X),
\end{split}
\eeq
where $Q$ is the doubly stochastic matrix $Q_{ij} = |\langle \varepsilon(X)_{i} | \varepsilon(Y)_{j} \rangle |^2$. Let $s$ be such that $\varepsilon(Y)_{j} \geq 0$ for $j \leq s$ and $\varepsilon(Y)_{j} < 0$ for $j > s$. Consider a term ``$m$'', $m \leq s$, from Eq.~\eqref{sup.as.disjoint.rank.sups}, i.e.,
\[
	\sup_{\substack{X \in \mathcal{C}, \\ \rank(X)=m}} \bigl\{   \vec{\varepsilon}(Y)^T \, Q \, \vec{\varepsilon}(X) - m  \bigr\},
\]
We claim that
\beq\label{maximum.of.eps.Q.eps}
	\vec{\varepsilon}(Y)^T \, Q \, \vec{\varepsilon}(X) \leq \vec{\varepsilon}(Y)^T (\underbrace{1,...,1}_{m-\text{times}},0,...,0)^T, \forall Q, \vec{\varepsilon}(X),
\eeq
is a tight upper bound. Consider
\beq\label{optimal.Q}
\begin{aligned}
        & \text{maximize} & & \vec{\varepsilon}(Y)^T \, Q \, \vec{\varepsilon}(X) \\
	& \text{subject to}
	& &\text{$Q$ doubly stochastic}.
\end{aligned}
\eeq
The optimization problem~\eqref{optimal.Q} is linear. It follows that the optimum is achieved at an extremal point. The doubly stochastic matrices form a polytope whose vertices are the permutation matrices (Birkhoff-von Neumann theorem). Hence, a solution $Q$ to~\eqref{optimal.Q} is a permutation matrix. An optimal choice is $Q = \id$ because $\vec{\varepsilon}(X)$ and $\vec{\varepsilon}(Y)$ are ordered descendingly, and
\[
	\langle x^{\downarrow}, y \rangle \leq \langle x^{\downarrow}, y^{\downarrow} \rangle
\]
for arbitrary vectors $x,y \in \AR^n$ (see Corollary II.4.4 in Bathia's book~\cite{Bathia1997}). Consequently, $Q = \id$, e.g., via
\[
	\ket{\varepsilon(X)_{j}} := \ket{\varepsilon(Y)_{j}}, \forall j,
\]
solves~\eqref{optimal.Q} independently of the specific values of $\vec{\varepsilon}(X)$ and $\vec{\varepsilon}(Y)$. To conclude the proof that Eq.~\eqref{maximum.of.eps.Q.eps} describes a tight upper bound, we have to solve
\beq\label{optimal.eps.X}
\begin{aligned}
        & \text{maximize} & & \vec{\varepsilon}(Y)^T \, \vec{\varepsilon}(X) \\
	& \text{subject to}
	& &X \geq 0, \; \| X \| \leq 1, \; \rank(X) = m.
\end{aligned}
\eeq
The constraints imply
\[
	(0,...,0)^T \leq \vec{\varepsilon}(X) \leq (\underbrace{1,...,1}_{m-\text{times}},0,...,0)^T
\]
(componentwise). As $m \leq s$, the l.h.s of Eq.~\eqref{maximum.of.eps.Q.eps} becomes maximal for the componentwise maximum of $\vec{\varepsilon}(X)$, i.e., for 
\beq\label{ub.case.m.leq.s}
	\vec{\varepsilon}(X) = (\underbrace{1,...,1}_{m-\text{times}},0,...,0)^T
\eeq
This proves that the upper bound in Eq.~\eqref{maximum.of.eps.Q.eps} is correct and tight. In case of $m > s$, non-zero choices of $\varepsilon(X)_{j}$, $s < j \leq m$, lead to negative contributions to the l.h.s of Eq.~\eqref{maximum.of.eps.Q.eps}. Hence, in case of $m > s$, the choice
\beq\label{ub.case.m.greater.s}
	\vec{\varepsilon}(X) = (\underbrace{1,...,1}_{s-\text{times}},0,...,0)^T.
\eeq
realizes the tight upper bound. Combining Eq.~\eqref{ub.case.m.leq.s} and Eq.~\eqref{ub.case.m.greater.s}, we arrive at
\begin{multline}
	\sup_{\substack{X \in \mathcal{C}, \\ \rank(X)=m}} \bigl\{   \vec{\varepsilon}(Y)^T \, Q \, \vec{\varepsilon}(X)\bigr\} - m \\
	=
	\left\{
	\begin{array}{ll}
  	\sum_{j=1}^m \bigl( \vec{\varepsilon}(Y)_{j} - 1 \bigr), &\text{ for } m \leq s \\
  	-(m-s) + \sum_{j=1}^s  \bigl( \vec{\varepsilon}(Y)_{j} - 1 \bigr), &\text{ for } m > s      
	\end{array} \right.
\end{multline}
To choose the optimal $m$ (recall Eq.~\eqref{sup.as.disjoint.rank.sups}), we note that $m \mapsto m+1$ is profitable as long as $\vec{\varepsilon}(Y)_{m} - 1 \geq 0$. Using the compact notation $a_{+} = \max\{ a,0 \}$, we conclude
\beq\label{rank.conjugate}
	\rank^*(Y) = \sum_{j=1}^n \bigl( \vec{\varepsilon}(Y)_{j} - 1 \bigr)_{+}.
\eeq
To determine $\rank^{**}(Z)$, we can copy and paste the Fazel-Hindi-Boyd arguments~\cite{FazelHindiBoyd2001}. We repeat them for the reader's convenience:
\beq\label{first.expression.for.rank.**}
	\rank^{**}(Z) = \sup_{Y=Y^T} \bigl\{ \tr(ZY) - \rank^*(Y) \bigr\}
\eeq
for all $Z \geq 0$ and $\| Z \| \leq 1$. Define
\beq
	\Omega := \bigl\{ \tr(ZY) - \rank^*(Y) \bigr\}.
\eeq
We consider the two cases $\| Y \| \leq 1$ and $\| Y \| > 1$, 
\beq\label{second.expression.rank**}
	\rank^{**}(Z) = \max\Bigl\{ \sup_{\substack{Y=Y^T, \\ \| Y \| \leq 1}} \Omega, \; \sup_{\substack{Y=Y^T, \\ \| Y \| > 1}}  \Omega  \Bigr\}.
\eeq
Assume $\| Y \| \leq 1$. Then, as a consequence of Eq.~\eqref{rank.conjugate}, $\rank^*(Y) = 0$, and therefore,
\beq\label{double.conjugate.of.rank.Y.norm.bounded}
	\sup_{\substack{Y=Y^T, \\ \| Y \| \leq 1}} \Omega = \sup_{\substack{Y=Y^T, \\ \| Y \| \leq 1}} \bigl\{ \tr(ZY) \bigr\}.
\eeq
By von Neumann's trace theorem~\cite{HornJohnson2},
\beq\label{von.Neumann.trace.theorem}
	\tr(ZY) \leq \vec{\varepsilon}(Z)^T \vec{\varepsilon}(Y).
\eeq
This upper bound can be achieved by choosing $Y$, such that
\[
	\ket{\varepsilon(Y)_{j}} := \ket{\varepsilon(Z)_{j}}, \forall j.
\]
Consequently, going back to Eq.~\eqref{double.conjugate.of.rank.Y.norm.bounded},
\beq
\begin{aligned}
        \sup \Omega = & \text{max} & & \vec{\varepsilon}(Z)^T \, \vec{\varepsilon}(Y) \\
	& \text{subject to}
	& & | \vec{\varepsilon}(Y)_{j} | \leq 1, \; \forall j.
\end{aligned}
\eeq
Since componentwise $0 \leq \vec{\varepsilon}(Z) \leq 1$, $\vec{\varepsilon}(Y) = (1,...,1)^T$ is the optimal choice. It follows that 
\beq\label{double.conjugate.of.rank.Y.norm.bounded.result}
	\sup_{\substack{Y=Y^T, \\ \| Y \| \leq 1}} \Omega = \sum_{j=1}^n \vec{\varepsilon}(Z)_{j} = \tr(Z).
\eeq
This concludes the discussion of $\| Y \| \leq 1$. Assume $\| Y \| > 1$. Note that $\rank^*(Y)$ is independent of our choice of the $Y$-eigenvectors $\ket{\varepsilon(Y)_{j}}$. Hence, in Eq.~\eqref{first.expression.for.rank.**}, we choose
\[
	\ket{\varepsilon(Y)_{j}} := \ket{\varepsilon(Z)_{j}}, \forall j,
\]
as before to reach the von Neumann-upper bound in Eq.~\eqref{von.Neumann.trace.theorem}. Thus, 
\beq
	\sup_{\substack{Y=Y^T, \\ \| Y \| > 1}} \Omega = \sup_{\substack{ \varepsilon(Y)_{1} \geq 1   }} \bigl\{ \vec{\varepsilon}(Z)^T \vec{\varepsilon}(Y) - \rank^*(Y) \bigr\}
\eeq
leading to
\beq
	\sup_{\substack{Y=Y^T, \\ \| Y \| > 1}} \Omega
	=	\sup_{\substack{ \varepsilon(Y)_{1} \geq 1   }} \sum_{j=1}^n (\varepsilon(Z)_{j} \varepsilon(Y)_{j}) - \sum_{j=1}^s \bigl( \varepsilon(Y)_{j} - 1 \bigr).
\eeq
Here, $s$ is chosen such that $\varepsilon(Y)_{j} \geq 1$ for $j \leq s$ and $\varepsilon(Y)_{j} < 1$ for $j > s$. As in the derivation by Fazel and coworkers~\cite{FazelHindiBoyd2001}, we continue by the addition and the subtraction of $\sum_{j=1}^n \varepsilon(Z)_{j}$:
\beq
\begin{split}	
	\sup_{\substack{Y=Y^T, \\ \| Y \| > 1}} \Omega
	&=	\sup_{\substack{ \varepsilon(Y)_{1} \geq 1   }} \sum_{j=1}^n (\varepsilon(Z)_{j} \varepsilon(Y)_{j}) - \sum_{j=1}^s \bigl( \varepsilon(Y)_{j} - 1 \bigr) \\  & \ \ \ \ \ \ \ \  - \sum_{j=1}^n \varepsilon(Z)_{j}  + \sum_{j=1}^n \varepsilon(Z)_{j}\\
	&=	\sup_{\substack{ \varepsilon(Y)_{1} \geq 1   }} \sum_{j=1}^s \bigl( \varepsilon(Y)_{j} -1 \bigr) \bigl( \varepsilon(Z)_{j} -1 \bigr) \\& \ \ \ \ \ \ \ \ \ \  \ +  \sum_{j=s+1}^n \bigl( \varepsilon(Y)_{j} -1 \bigr)\varepsilon(Z)_{j}   + \sum_{j=1}^n \varepsilon(Z)_{j}.
\end{split}
\eeq
In this last expression, the first sum is negative semidefinite because $\| Z \| \leq 1$, and the second sum is negative semidefinite because by definition of $s$, $\varepsilon(Y)_{j} \leq 1$ for all $j > s$. Therefore,
\beq\label{double.conjugate.of.rank.Y.norm.UNbounded.result}
	\sup_{\substack{Y=Y^T, \\ \| Y \| > 1}} \Omega \leq \sum_{j=1}^n \varepsilon(Z)_{j} = \tr(Z).
\eeq
Hence, using $Y$ with $\| Y \| > 1$ brings no advantage (compare Eq.~\eqref{double.conjugate.of.rank.Y.norm.bounded.result} and Eq.~\eqref{double.conjugate.of.rank.Y.norm.UNbounded.result}). Going back to Eq.~\eqref{second.expression.rank**}, we conclude
\beq
	\rank^{**}(Z) = \tr(Z),
\eeq
i.e., the convex envelope of the matrix rank function over the set $S^+ \cap B_{\| . \| \leq 1}$ is the matrix trace.

\subsection{Numerical experiments}

To perform the numerical experiments, we need to sample explicit states and measurements. We proceeded by choosing pure states from the Haar measure and by rotating a reference projective, non-degenerate measurement according to the Haar measure. In Algorithm~1 in the main text we need to compare the ranks of matrices. In principle, the rank of a matrix is equal to the number of its non-zero singular values. However, due to small numerical fluctuations in the solutions $G$, this definition is too strict. Rather, one should tolerate small variations by setting to zero singular values that are very small. We need to compare the rank of $G$ with the rank of $\mathcal{D}$. We proceed by defining a threshold $\tau := 10^{-4}$ and
\beq
	\check{s} := (s_{\rank{\mathcal{D}+1}}, ..., s_{N}),
\eeq
with $s_{j}$ denoting the singular values of $G_{\mathrm{st}}$ (sorted descendingly). Then, we choose the following criterion to decide whether or not the ranks of $G$ and $\mathcal{D}$ agree:
\beq
	\rank(G) \approx_{\tau} \rank(\mathcal{D}) \; :\Leftrightarrow \; \| \check{s} \|_{2} \leq \tau.
\eeq



\end{document}